\documentclass[a4paper]{jpconf}
\begin{document}

\title{Supermassive binary black hole mergers}
\author{L\'{a}szl\'{o} \'{A}. Gergely$^{1,2,3\star }$ and Peter L. Biermann$%
^{4,5,6\ddag }$}

\address{
$^{1}$Department of Theoretical Physics, University of Szeged, Tisza Lajos krt 84-86, 
Szeged 6720, Hungary\\
$^{2}$Department of Experimental Physics, University of Szeged, D\'{o}m t\'{e}r 9, 
Szeged 6720, Hungary\\
$^{3}$Department of Applied Science, London South Bank University, 103 Borough Road, 
London SE1 0AA, UK\\
$^{4}$Max Planck Institute for Radioastronomy, Bonn, Germany\\
$^{5}$Department of Physics and Astronomy, University of Bonn, Germany\\
$^{6}$Department of Physics and Astronomy, University of Alabama,
Tuscaloosa, AL, USA }

\ead{$^\star$ gergely@physx.u-szeged.hu\qquad
     $^\ddag$ plbiermann@mpifr-bonn.mpg.de}

\begin{abstract}
When galaxies collide, dynamical friction drives their central supermassive
black holes close enought to each other such that gravitational radiation
becomes the leading dissipative effect. Gravitational radiation takes away
energy, momentum and angular momentum from the compact binary, such that the
black holes finally merge. In the process, the spin of the dominant black
hole is reoriented. On observational level, the spins are directly related
to the jets, which can be seen at radio frequencies. Images of the X-shaped
radio galaxies together with evidence on the age of the jets illustrate that
the jets are reoriented, a phenomenon known as spin-flip. Based on the
galaxy luminosity statistics we argue here that the typical galaxy
encounters involve mass ratios between $1:3$ to $1:30$ for the central black
holes. Based on the spin-orbit precession and gravitational radiation we
also argue that for this typical mass ratio in the inspiral phase of the
merger the initially dominant orbital angular momentum will become smaller
than the spin, which will be reoriented. We prove here that the spin-flip
phenomenon typically occurs already in the inspiral phase, and as such is
describable by post-Newtonian techniques.
\end{abstract}

\section{Introduction}

Einstein's general relativity is a rigorous theory of gravitation reducing
to Newton's theory under the proper limit of slow motions and weak gravity.
The rigorousness comes together with incredible geometrical elegance, based
however on a more complicated mathematical description, according to which
even the two-body problem has no exact solution. Two bodies orbiting each
other driven by their mutual gravitational attraction lack the usual
conserved dynamical quantities. Indeed, the energy, momentum and angular
momentum all decrease in the 2.5th order of the perturbations due to the
escaping gravitational radiation, a novel prediction of general relativity.
The loss of energy leads to the reduction of the orbital period, the loss of
orbital angular momentum leads to the circularization of the
quasi-elliptical orbits, while the loss of momentum can induce in certain
extreme cases a strong recoil effect of the merged black hole (Bruegmann
2007, Gonzalez et al. 2007a, b), leading even to its ejection from the host
galaxy.

Although gravitational waves were not detected directly yet, there is
overwhelming indirect evidence for their existence, the first of them being
the reduction of orbital period of the Hulse-Taylor pulsar in precisely the
way predicted by general relativity. The first generation of Earth-based
interferometric detection devices (LIGO, VIRGO) built to detect
gravitational waves emitted by relatively close sources are already
operational and the LISA space mission is expected to further increase not
only the likelihood to directly detect gravitational waves, but also to
deepen our understanding of astrophysical phenomena occurring in the
gravitational wave sources. Compact binaries in the latest stages of their
merge are among the most important sources for gravitational waves.

After the initial growth of galaxies, their evolution is governed by
mergers. As most galaxies have a central black hole, these will also merge.
(see Rottmann 2001; Zier \& Biermann 2001, 2002; Biermann et al. 2000;
Merritt \& Ekers 2002; Merritt 2003; Gopal-Krishna et al. 2000, 2003, 2004,
2006; Gopal-Krishna \& Wiita 2006; Zier 2005, 2006a, 2006b). At the
beginning of this process the galaxies get distorted by their mutual
interaction.

The interaction of the central black holes with the already merged stellar
environment generates a dynamical friction. This is the leading dissipative
effect when the separation of the black holes is between a few parsecs and
one hundredth of a parsec. In this process, some of the orbital angular
momentum of the binary black hole system is transferred to the stellar
environment, such that the stellar population at the poles of the system
tends to be ejected and a torus is formed (Zier \& Biermann 2001, Zier
2006a). There had been a major worry, that the two black holes stall in
their approach to each other (Valtonen 1996, Yu 2003, Merritt 2003, 2005,
Milosavljevi{\'c} \& Merritt 2003a, b, Makino \& Funato 2004, Berczik et al.
2005, 2006, Matsubashi et al. 2007) and they will not get to such an
approach, which will let them to significant emission of gravitational
waves. According to these worries the loss-cone mechanism for feeding stars
into orbits that intersect the binary black holes is too slow. However,
recently Zier (2006a) has demonstrated that direct interaction with the
surrounding stars slightly further outside speeds up the process. New work
by Merritt, Mikkola \& Szell (2007) is consistent with Zier (2006a). On the
other side, relaxation processes due to cloud/star-star interactions are
rather strong, as shown by Alexander (2007), using observations of our
galaxy. These interactions repopulate the stellar orbits in the center of
the galaxy. Therefore is very likely that no stalling occurs and the compact
binary arrives in the regime, where the emission of gravitational radiation
has an important impact on the dynamics.

Galactic black holes spin fast. Along their spin axis, a jet is emitted.
When two galaxies merge, their mass is rarely comparable, and in consequence
the dominance of one spin and the formation of one pair of jets is typical.
If the spin axis is reoriented during the merger, a new jet will be formed.
This process can be seen in X-shaped radio galaxies (Rottmann 2001, Chirvasa
2001, Biermann et al. 2000, Merritt and Ekers 2002). These contain two pairs
of jets, typically at apparent angle which is less than 30 degrees
(therefore the real angle is about 45 degrees). One pair of jets has a steep
radio spectrum, thus it has not recently been resupplied energetically, it
is an old pair of jets. The other pair of jets has a relatively flat radio
spectrum (this is the new jet) (Rottmann 2001). The observations thus
support the spin-flip model.

Another key observation is, that the analysis of the radio spectrum suggests
that the spin of the black hole both before and after the merger is high,
more than $95~\%$ of the maximally allowed value (Falcke et al. 1995,
Biermann et al. 1995, Falcke and Biermann 1995a, 1995b, 1999, Donea \&
Biermann 1996, Mahadevan 1998, Gopal-Krishna et al. 2004). This is a major
constraint any model explaining the spin-flip phenomenon should obey.

The potential source for the spin-flip is the existence of a much higher
orbital angular momentum of the compact binary then the spin at the
beginning of the merger, combined with the situation of no orbital angular
momentum left after the merger occurred. Obviously in the process of loosing
the orbital angular momentum, the spin could have absorbed some of it and as
its magnitude can not increase further to much, simply get reoriented.

The gravitational radiation dominated regime of the merging process can be
divided into three phases. The inspiral phase can be well described by
analytical (post-Newtonian, PN) techniques. The PN formalism breaks however
down close to the innermost stable orbit (ISO) and only a numerical
treatment of the plunge phase is possible. Finally, in the ringdown phase
the already merged object will loose all characteristics which carry the
imprints of its history, with the exception of mass, spin and possibly
(although unlikely) also electric charge.

Previous numerical works aimed to explain the spin-flip phenomenon as
occurring during the plunge, have succeeded in certain particular mass and
initial spin configurations (Campanelli et al. 2007a, b, and references
therein), however the merged black hole not always has a high spin as
required by observations.

We present here a new approach (Gergely \& Biermann 2007), based on the

(a) the galaxy luminosity statistics, which allows us to show that the
typical mass ratio of the two supermassive black holes is $\eta=1/3$ to $%
\eta=1/30$.

(b) the analysis of the spin evolution based on the leading order
conservative dynamical effect, the spin-orbit interaction; and the leading
order dissipative effect, gravitational radiation. This formalism has been
worked out a long time ago (Apostolatos et al. 1994), however we exploit
here the consequences of the fact that in the typical mass range there are
two small parameters in the formalism.

Beside the PN parameter $\varepsilon$, which increases from about $10^{-3}$
at the time when gravitational radiation overtakes the dynamical friction as
the leading dissipative effect to at about $0.1$, the second small
parameter, the mass ratio $\eta$ stays constant. The interplay of these two
small parameters guarantees that for the typical mass ratio range at the
beginning of the inspiral phase the orbital angular momentum dominates over
the spin; however at the end of the inspiral there is not much angular
momentum left, the spin is the dominant angular momentum. The spin flip has
already occurred during the inspiral and will not change significantly
during the plunge or ringdown.

\section{Typical mass ranges}

Lauer et al. (2006) presents the mass distribution of galactic central black
holes, basically confirming earlier work (Press \& Schechter 1974). This is
consistent with a recent observational survey (Ferrarese at al. 2006) and
the discussion of Wilson \& Colbert (1995). The black hole mass distribution 
$\Phi _{BH}(M_{BH})$ can be described as a broken powerlaw, from about $%
m_{a}\simeq 3\times 10^{6}$ solar masses ($M_{\odot }$) to about $%
m_{b}\simeq 3\times 10^{9}$ $M_{\odot }$, with a break near $m_{\star
}\simeq 10^{8}$ $M_{\odot }$. The values of $m_{a},~m_{b}$ and $m_{\ast }$
imply that we have two mass ranges of a factor of $30$ each. The masses
above $10^{8}$ $M_{\odot }$ are rapidly becoming rare with higher mass.

The mass of the central massive black holes scales with the total mass of a
galaxy (the dark matter). As argued by Zier (2006a) the approach of the two
black holes does not stall, and each merger of two massive galaxies will
also lead to the merger of the two central black holes. Observational
evidence suggests that black holes merge on the rather short time scales of
active galactic nuclei. We use the merger rate of galaxies as closely
equivalent to the merger rate of the central black holes.

The statistics of the mergers arises from the integral giving the number of
mergers $N(\eta )$ per volume and time, for any mass ratio $\eta $ defined
to be smaller than unity. This integral is given by the product of the mass
distribution of the first black hole with the mass distribution of the
second black hole multiplied by a merger rate $F$. Due to its larger
cross-section, the more massive black hole will dominate the merger rate $F$%
, so that it can be approximated as a function of $\eta ^{-1}m$ alone and
the dependence on $\eta $ can be taken as a power law behavior with $F\sim
\eta ^{-\xi }$ with $\xi =1/2$ (Gergely \& Biermann 2007). As the black hole
mass distribution has a break at $\eta ^{-1}=30$, we use $\Phi _{BH}(m)\sim
m^{-\alpha }$ for the first mass range, and $\Phi _{BH}(m)\sim m^{-\beta }$
for the second. For a given $\eta $ in the range $\eta ^{-1}$ from $1$ to $%
30 $ the number of mergers scales as

\begin{eqnarray}
N(\eta ) &\sim &\int_{m_{a}}^{\eta m_{\star }}\left( \frac{m}{m_{\star }}%
\right) ^{-\alpha }\left( \frac{m}{\eta m_{\star }}\right) ^{-\alpha }\left( 
\frac{m}{\eta m_{\star }}\right) ^{\xi }dm  \nonumber \\
&+&\int_{\eta m_{\star }}^{m_{\star }}\left( \frac{m}{m_{\star }}\right)
^{-\alpha }\left( \frac{m}{\eta m_{\star }}\right) ^{-\beta }\left( \frac{m}{%
\eta m_{\star }}\right) ^{\xi }dm  \nonumber \\
&+&\int_{m_{\star }}^{\eta m_{b}}\left( \frac{m}{m_{\star }}\right) ^{-\beta
}\left( \frac{m}{\eta m_{\star }}\right) ^{-\beta }\left( \frac{m}{\eta
m_{\star }}\right) ^{\xi }dm
\end{eqnarray}%
while for $\eta ^{-1}$ above $30$:

\begin{equation}
N(\eta )\sim \int_{m_{a}}^{\eta m_{b}}\left( \frac{m}{m_{\star }}\right)
^{-\alpha }\left( \frac{m}{\eta m_{\star }}\right) ^{-\beta }\left( \frac{m}{%
\eta m_{\star }}\right) ^{\xi }dm~.
\end{equation}%
(We have dropped the most rare encounters of black holes with extremely high
masses between $\eta ^{-1}m_{b}$ and $m_{b}$.)

According to the models presented in Lauer et al (2006) we take $\alpha =1$
and $\beta =3$. Then the above integrands are monotonically decreasing
functions, the integrals being dominated by the lower limits. The four terms
then scale with $\eta $ as ${\eta }^{\alpha -\xi }$, $\eta ^{1-\alpha }$, ${%
\eta }^{\beta -\xi }$, ${\eta }^{\beta -\xi }$. The first term contains
small galaxies merging with small galaxies for which the cross section is
low. Even so, one can see that the more extreme mass ratios are more common
as the distribution of the number of mergers in the mass ratio range $1:30$
to $1:3$ versus $1:3$ to $1:1$ is approximately $5$. For the second term
describing massive galaxies merging with smaller galaxies this ratio of
mergers in the two mass ratio ranges is about $14$. The third term is almost
negligible, and the fourth term adds cases to the second term with more
extreme mass ratios, above $1:30$, and so emphasizes the large mass ratio
range. We conclude that the most common mass ratio range is $1:3$ to $1:30$.

\section{The spin-flip}

Under the combined effect of spin-orbit precession (at 1.5 PN orders) and
the gravitational radiation (at 2.5 PN orders), the direction and magnitude
of the dominant spin and orbital angular momentum evolve as (Apostolatos et
al. 1994):%
\begin{eqnarray}
\dot{S}_{1} &=&0~,\qquad \mathbf{\skew{0}{\dot}{\hat{S}}_{1}}={\frac{2G}{%
c^{2}r^{3}}}\mathbf{J}\times \mathbf{\hat{S}_{1}}\ ,  \nonumber \\
\dot{L} &=&-\frac{32G\mu ^{2}}{5r}\left( \frac{Gm}{c^{2}r}\right)
^{5/2}~,\qquad \mathbf{\skew{0}{\dot}{\hat{L}}}={\frac{2G}{c^{2}r^{3}}}%
\mathbf{J\times \hat{L}}~.  \label{dynamics}
\end{eqnarray}%
Due to the spin-orbit interaction (also discussed in Apostolatos et al.
1994, Kidder 1995, Ryan 1996, Rieth \& Sch\"{a}fer 1997, Gergely at al.
1998a, 1998b, 1998c, O'Connell 2004), both $\mathbf{L}$ and $\mathbf{S}_{%
\mathbf{1}}$ undergo a precessional motion about $\mathbf{J}$. The spin-spin
(Kidder et al. 2003, Kidder 1995, Apostolatos 1995, Apostolatos 1996,
Gergely 2000a, 2000b), mass quadrupolar (Poisson 1998, Gergely \& Keresztes
2003), possible magnetic dipolar (Ioka \& Taniguchi 2000, Vas\'{u}th et al.
2003), self-spin (Mik\'{o}czi et al. 2005) and higher order spin-orbit
effects (Faye et al. 2006, Blanchet et al. 2006) can slightly modulate this
process.

The total angular momentum $\mathbf{J=S}_{1}+\mathbf{L}$ is changed by the
emitted gravitational radiation as $\mathbf{\dot{J}=}\dot{L}\mathbf{\hat{L}}$%
, thus%
\begin{equation}
\dot{J}=\dot{L}\left( \mathbf{\hat{L}}\cdot \mathbf{\hat{J}}\right) \mathbf{%
~,\qquad \skew{0}{\dot}{\hat{J}}}=\frac{\dot{L}}{J}\left[ \mathbf{\hat{L}}%
-\left( \mathbf{\hat{L}}\cdot \mathbf{\hat{J}}\right) \mathbf{\hat{J}}\right]
~.  \label{Jdot}
\end{equation}%
As the time scale of the gravitational radiation driven orbital shrinking is
much higher than the precession time-scale, the first term in $\mathbf{%
\skew{0}{\dot}{\hat{J}}}$ is averaged out over one precession. Therefore in
these cases (known as simple precession), in an averaged sense $\mathbf{J}$
changes only along itself, in other words its direction is conserved. The
angles $\lambda =\cos ^{-1}\left( \mathbf{\hat{L}\cdot \hat{J}}\right) $ and 
$\sigma =\cos ^{-1}\left( \mathbf{\hat{S}}_{\mathbf{1}}\mathbf{\cdot \hat{J}}%
\right) $ evolve as%
\begin{equation}
\dot{\lambda}=-\frac{\dot{L}}{J}\sin \lambda >0~,\qquad \dot{\sigma}=\frac{%
\dot{L}}{J}\sin \lambda <0~.  \label{betadot}
\end{equation}%
Therefore during the inspiral $\mathbf{J}$ shrinks but keeps its direction; $%
\mathbf{L}$ decreases and tilts away from $\mathbf{J}$; while $\mathbf{S}%
_{1} $ remains unchanged in magnitude and tilts towards $\mathbf{J}$. As a
result, the orbital angular momentum slowly turns away from $\mathbf{J}$,
while $\mathbf{S_{1}}$ slowly approaches the direction of $\mathbf{J}$.

Let us focus on the magnitudes of these vectors. For this we estimate $%
S_{1}/L\approx \varepsilon ^{1/2}\eta ^{-1}$~(Gergely \& Biermann 2007). In
the domain $\eta =1/3$ to $1/30$ as the PN parameter evolves from $%
\varepsilon =10^{-3}$ to $10^{-1}$, the situation changes from $L$ being
dominant over $S_{1}$ to $S_{1}$ dominating over $L\,\ $nearby ICO.
Therefore we conclude that in these typical cases the spin has tilted close
to the conserved direction $\mathbf{\skew{0}{\dot}{\hat{J}}}$, while not
much orbital momentum has left. In other words, the spin-flip has occurred.

\section{Concluding Remarks}

The jets of X-shaped radio galaxies, which represent the dominant spin
before the merger and the resulting spin after the merger represent the spin
directions in the initial and final configuration. We have shown here that
the typical mass range when galactic black holes merge is $1:3$ to $1:30$.
In this range the combined effects of the spin-orbit precession and
gravitational radiation result in

(i.) a shrinking of the orbital angular momentum, from a dominant value at
the beginning of the inspiral to a sub-dominant value at the end of the PN
regime. This means that in later stages of the merger the orbital angular
momentum cannot influence too much the spin evolution.

(ii.) a precession of the spin vector during this process, which becomes
faster as the black holes approach each other. This can be thought as a
super-wind sweeping away the base of the old jet, in accordance with
observations (Gopal-Krishna et al. 2003, 2006, Gopal-Krishna \& Wiita 2006).

(iii.) a reorientation of the spin vector and of the jet towards the
direction towards which the initial orbital angular momentum was pointing.
This is occurring in the inspiral phase already and can be discussed through
analytical methods.

(iv.) as the magnitude of the spin is unchanged by the whole process, its
value can stay high if it was high before the merger.

In Gergely \& Biermann (2007) we show that the super-disk radiogalaxies are
plausible candidates for undergoing a spin-flip soon, and afterwards such
galaxies are prime candidates to produce very energetic neutrinos and other
very high energy particles. Preliminary investigations show that the time
scale is rapid enough to produce the clear flip seen in the x-shaped radio
galaxies, but more rigurous calculations have to be done in order to
quantitatively test this mechanism.

\section{Acknowledgements}

We wish to acknowledge intense discussions with Gopal-Krishna and Christian
Zier. PLB wishes to acknowledge further discussions with J. Barnes, B. Br{%
\"{u}}gmann, and G. Sch{\"{a}}fer. L\'{A}G was supported by OTKA grants no.
46939, 69036 and the J\'{a}nos Bolyai Scholarship of the Hungarian Academy
of Sciences. Support for PLB is coming from the AUGER membership and theory
grant 05 CU 5PD 1/2 via DESY/BMBF. The collaboration between the University
of Szeged and the University of Bonn is via an EU Sokrates/Erasmus contract.
L\'{A}G wishes to thank the organizers of the Amaldi7 meeting for financial
support.

\end{document}